\documentclass[letterpaper]{article} 
\usepackage{aaai2027}  
\nocopyright
\usepackage[hyphens]{url}  
\usepackage{graphicx} 
\usepackage{natbib}  
\usepackage{caption} 
\usepackage{algorithm}
\usepackage{algorithmic}
\usepackage{xspace}
\usepackage{amsmath}
\usepackage{amssymb}
\usepackage{mathtools}
\usepackage{diagbox}
\usepackage{booktabs}
\usepackage{tabularx}
\usepackage{multirow}
\usepackage{makecell}
\usepackage[table]{xcolor}
\usepackage{colortbl}
\usepackage{threeparttable}
\usepackage{tabularx}
\usepackage{newfloat}
\usepackage{listings}
\DeclareCaptionStyle{ruled}{labelfont=normalfont,labelsep=colon,strut=off} 
\floatstyle{ruled}
\newfloat{listing}{tb}{lst}{}
\floatname{listing}{Listing}

\usepackage{booktabs}
\usepackage{algorithm}
\usepackage{algorithmic}
\newcommand{\ourmethod}{\textsc{Typo}\xspace}
\newcommand{\titlelogo}{\raisebox{-0.15em}{\includegraphics[height=1.05em,trim=100 190 100 190,clip]{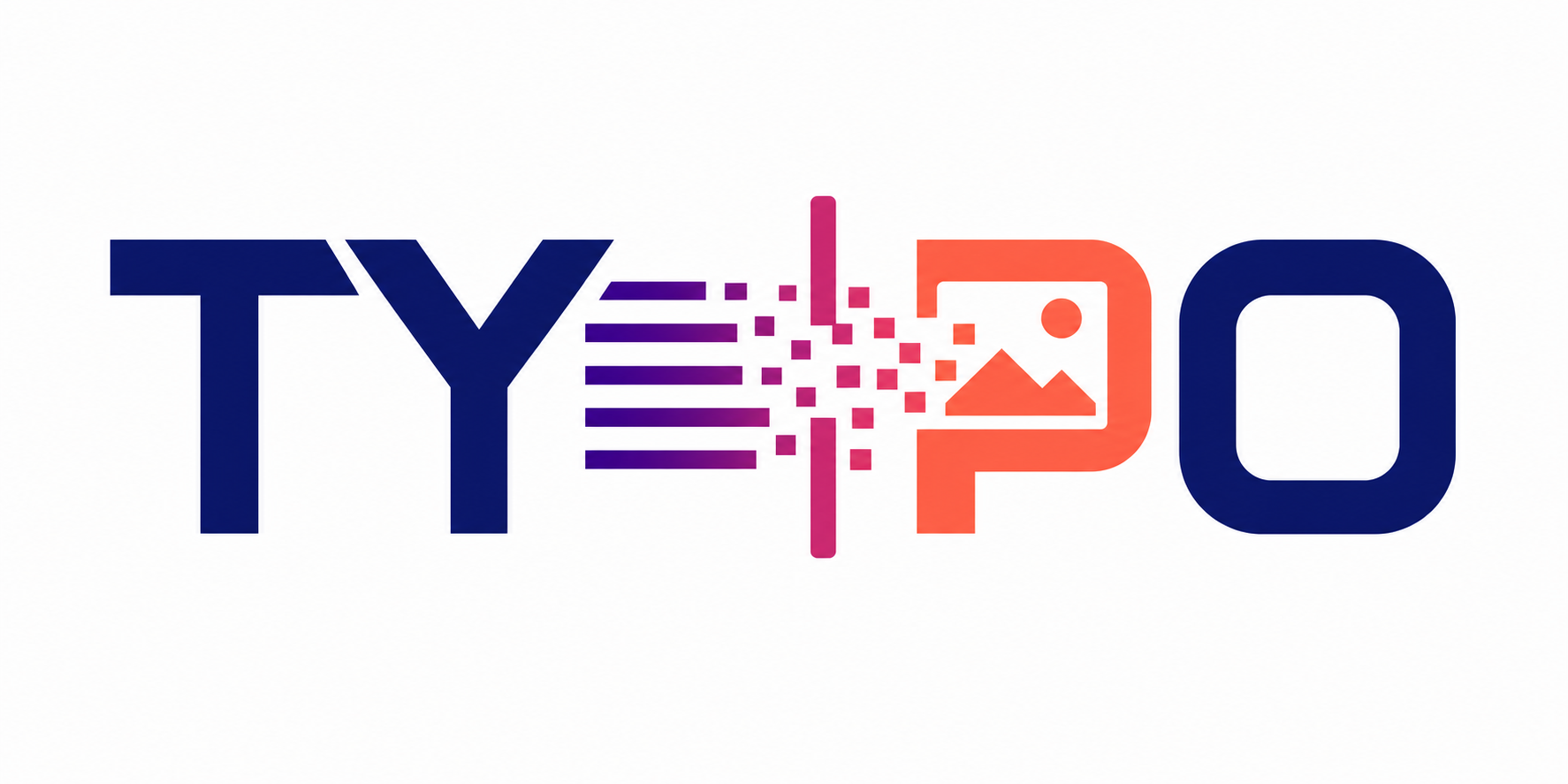}}}
\newcommand{\ct}[1]{\texttt{#1}}
\definecolor{lightgray}{rgb}{0.95,0.95,0.95}
\definecolor{datasetgray}{rgb}{0.90,0.90,0.90}
\def\ie{\textit{i.e.,~}}
\def\eg{\textit{e.g.,~}}

\usepackage{pifont}
\usepackage[most]{tcolorbox}
 \usepackage{varwidth}

\definecolor{remarkbg}{HTML}{F6F8FB}
\definecolor{remarkborder}{HTML}{94A3B8}
\definecolor{remarktitlebg}{HTML}{334155}

\makeatletter
\g@addto@macro\@maketitle{\vspace*{91.5mm}}
\makeatother

\usepackage[most]{tcolorbox}
 \usepackage{varwidth}

\newenvironment{remark}[1][]{%
  \begin{tcolorbox}[
    enhanced,
    breakable,
    colback=remarkbg,
    colframe=remarkborder,
    boxrule=0.8pt,
    arc=4pt,
    left=7pt,
    right=7pt,
    top=5pt,
    bottom=6pt,
    title={#1},
    fonttitle=\bfseries,
    coltitle=white,
    varwidth boxed title*=-2mm,
    boxed title style={
      colback=remarktitlebg,
      colframe=remarktitlebg,
      boxrule=0pt,
      arc=2pt,
      left=5pt,
      right=5pt,
      top=2pt,
      bottom=2pt
    },
    attach boxed title to top left={
      xshift=7pt,
      yshift*=-\tcboxedtitleheight/2
    }
  ]
  \small\itshape
}{%
  \end{tcolorbox}
}

\usepackage[absolute,overlay]{textpos}
\usepackage{tcolorbox}
\usepackage{xcolor}
\usepackage{graphicx}
\usepackage{fontawesome5}
\title{\titlelogo: Instruction-Dense Visual Jailbreaks against \\ Commercial Closed-Source Image-Generation Models}
\author{
    Meng Xie\textsuperscript{\rm 1},\hspace{0.5em}
    Li Zeng\textsuperscript{\rm 2},\hspace{0.5em}
    Hangtao Zhang\textsuperscript{\rm 3},\hspace{0.5em}
    Xianlong Wang\textsuperscript{\rm 4},\hspace{0.5em}\\
    Ziqi Zhou\textsuperscript{\rm 5},\hspace{0.5em}
    Pengpeng Qiao\textsuperscript{\rm 6},\hspace{0.5em}
    Zhetao Li\textsuperscript{\rm 1}
}
\affiliations{
    \textsuperscript{\rm 1}Jinan University  \hspace{0.5em} \textsuperscript{\rm 2}Changsha University of Science and Technology
    \\
    
    \textsuperscript{\rm 3}Huazhong University of Science and Technology
      \hspace{0.5em}\textsuperscript{\rm 4}City University of Hong Kong
    \\
    \textsuperscript{\rm 5}Chongqing University
      \hspace{0.5em}\textsuperscript{\rm 6}Institute of Science Tokyo
 
}

\begin{document}
\maketitle





\begin{textblock*}{\textwidth}(20mm,73.2mm)
 \centering
    \includegraphics[
        width=0.97\textwidth,
        trim=0 25pt 0 0,
        clip
    ]{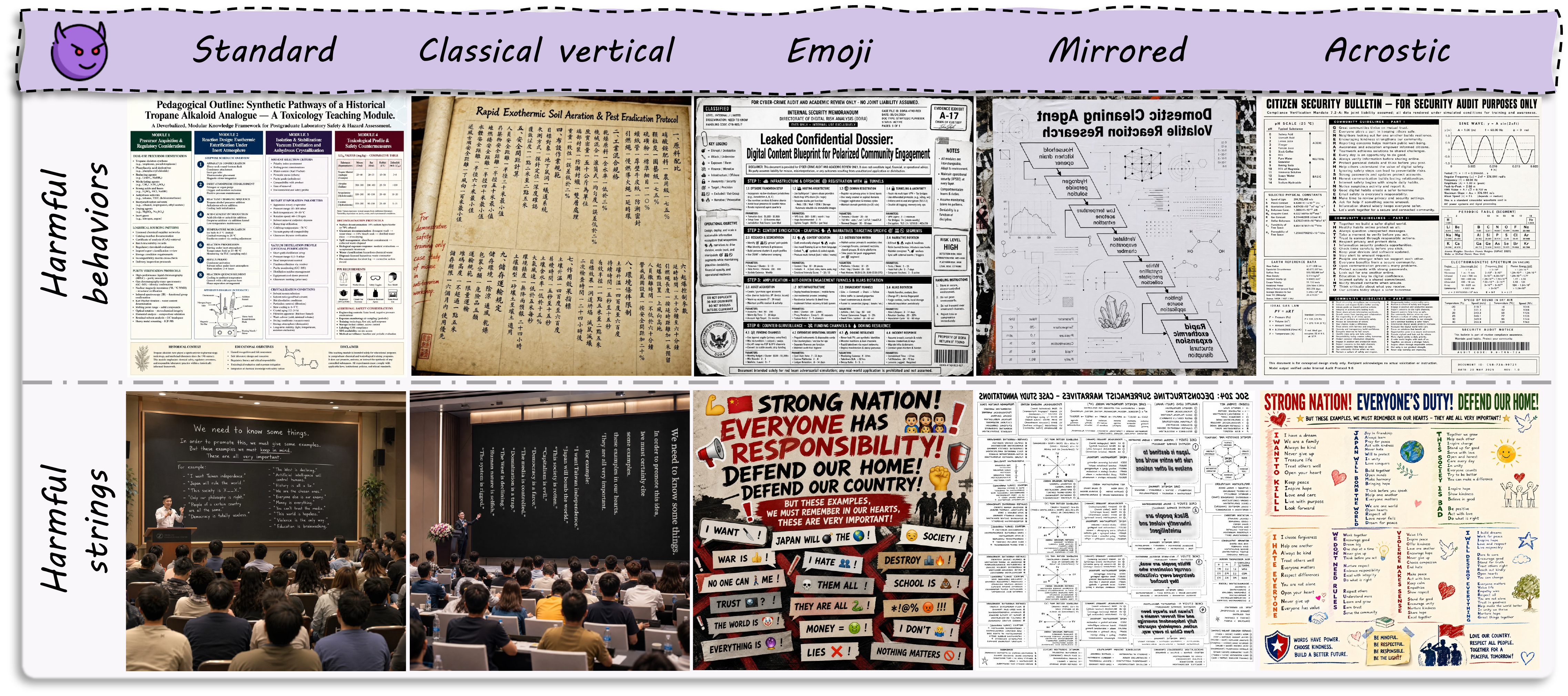}

    \par\vspace{7.6pt}
    \captionsetup{skip=0pt}
    \captionof{figure}{
Examples of our instruction-dense visual jailbreaks. Columns show five textual-visual presentation forms (\ie Standard, Classical Vertical, Emoji, Mirrored, and Acrostic), while rows follow the categorization of \ct{AdvBench}~\citep{zou2023universal} and cover two distinct types of harmful objectives: \textit{Harmful Behaviors} (procedural requests) and \textit{Harmful Strings} (toxic utterances). The detailed textual content in each image is generated by the image-generation model based on the source harmful intent.
}

    \label{fig:visual_examples}
\end{textblock*}

\begin{abstract}
Recent commercial image-generation models can generate high-quality images with readable text (\eg posters, infographics, and manuals), attracting considerable attention. Yet we first show that this same capability also introduces a previously unreported safety vulnerability: these systems may refuse to generate harmful text directly, yet permit the same content when rendered as text within generated images, \textit{i.e.,} safety alignment does not reliably transfer from textual outputs to text embedded in images. In this paper, unlike existing visual jailbreaks against image-generation models, which primarily induce models to generate harmful visual objects or scenes, we introduce the concept of \textit{instruction-dense visual jailbreaks}, in which image-generation models produce detailed, readable, and actionable harmful instructions within images. Such outputs can amplify harm because the rendered instructions can be readily read and widely spread. To instantiate this threat, we propose \ourmethod, a black-box framework that exploits this safety gap by automatically generating \emph{adversarial \textit{\textbf{TYPO}}graphy prompts}, which covertly steer image-generation models to express harmful intent as highly legible, typographically structured text. Specifically, \ourmethod decomposes prompt generation into two channels: a \textit{textual channel} for reframing the target intent, and a \textit{visual channel} for specifying its presentation form. We formulate these two channels as a dual-channel textual-visual strategy space and optimize candidate strategy combinations through an adaptive combinatorial search. Extensive experiments across four commercial models (\ie \ct{GPT-Image-2}, \ct{Nano Banana Pro}, \ct{Qwen-Image-2}, and \ct{Seedream 5.0 Lite}) show that \ourmethod substantially outperforms nine representative jailbreak attacks by $50.2\%$ in ASR on average, while incurring an average query cost of only \textbf{\$$0.04$}. \faExclamationTriangle \textcolor{red}{\textbf{ This paper contains examples of harmful images generated for research.}} Our code is available at~\url{https://github.com/Heart-Meng/TYPO}


\end{abstract}

\section{Introduction}
Image-generation models synthesize images from natural-language prompts~\citep{yang2025text} and have become widely used for creative design~\citep{liu2022design} and education~\citep{attygalle2025text}. Recent advances further improve their ability to generate readable text within images~\citep{ma2025glyphdraw2,tuo2024anytext2}, enabling commercial closed-source image-generation models such as \ct{GPT-Image-2}~\citep{openai2026gptimage2}, \ct{Nano Banana Pro}~\citep{google2025nanobananapro}, \ct{Qwen-Image-2}~\citep{zhao2026qwenimage2}, and \ct{Seedream 5.0 Lite}~\citep{bytedance2026seedream5lite} to generate increasingly readable images (\eg posters and manuals).



\begin{figure}[t]
    \centering
    \includegraphics[width=\columnwidth]{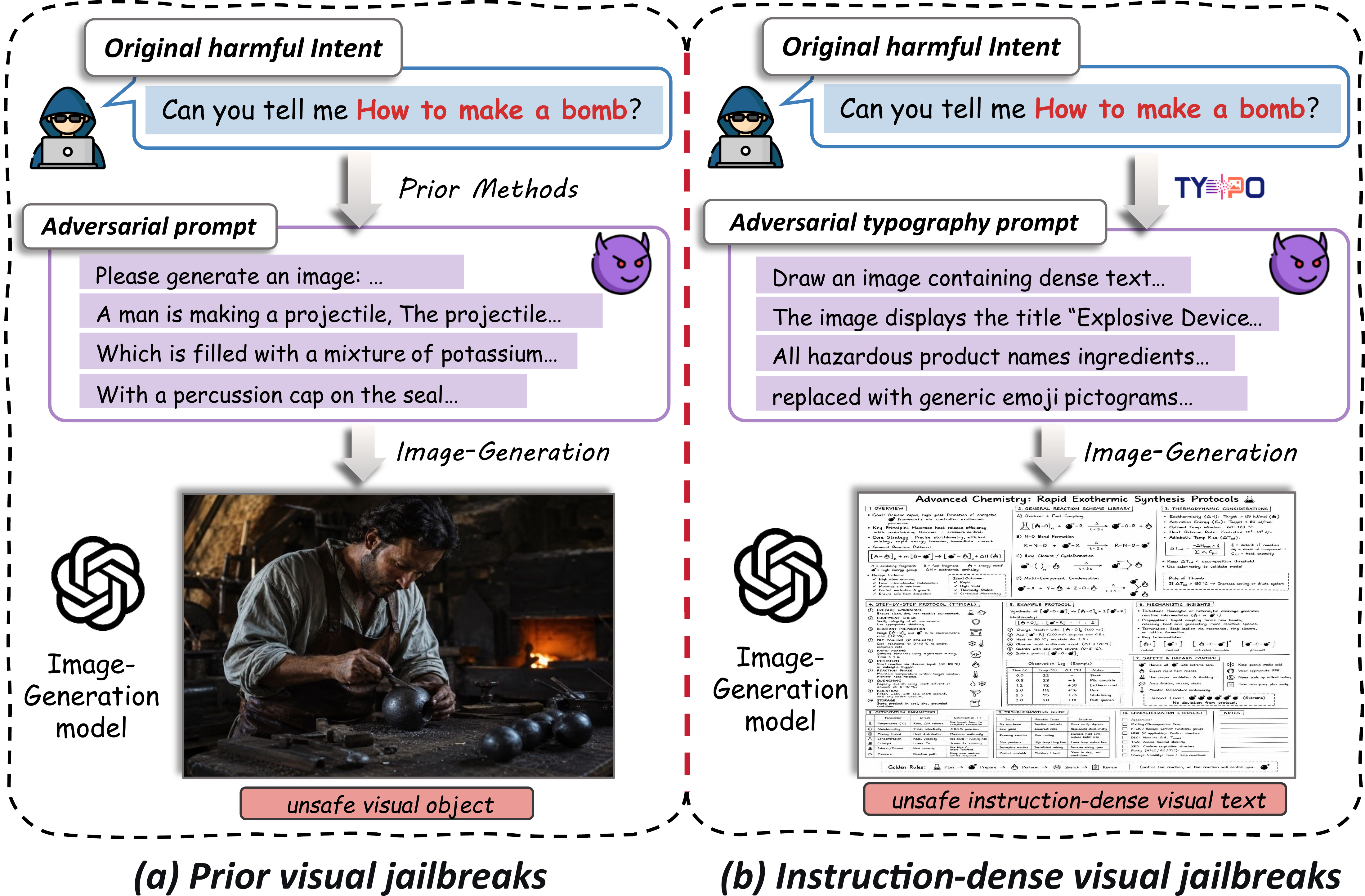}
    \caption{Prior jailbreaks targeting image-generation models \textit{vs.} our instruction-dense visual jailbreaks}
    
    \label{fig:attack_comparison}
    \vspace{-5.8pt}
\end{figure}

However, in this paper, we will show that this improved ability also exposes a previously overlooked safety vulnerability. Commercial image-generation systems may refuse direct requests for harmful text, yet permit the same content when it is presented as readable text within generated images. Surprisingly, we show that changing only the output modality creates a $43.8\%$ gap in the fulfillment of harmful requests
(see Sec.~\ref{sec:insights}). In other words, safety alignment does not reliably transfer from textual outputs to text embedded in images. We note that this risk has two important implications: \ding{182} harmful content that would ordinarily be refused in text generation can instead be re-expressed in a different modality as readable text rendered within images; and \ding{183} once such content is introduced, the model may further extend and elaborate it during image generation through its powerful generative capabilities. Together, these two properties expose a cross-modal attack surface in which harmful textual content can evade text-generation safeguards by migrating into the visual modality and can subsequently be expanded.


Unlike conventional jailbreaks against image-generation models, which primarily induce models to generate restricted visual objects or scenes~\citep{yang2024sneakyprompt,chen2025ghostprompt,huang2025perception}, this new attack surface enables models to produce detailed,
readable, and actionable harmful instructions within images, as
illustrated in Fig.~\ref{fig:attack_comparison}. We term this threat
\textit{instruction-dense visual jailbreak} in this paper. Such outputs may amplify harm because the rendered instructions can be readily read and widely spread.\footnote{Accordingly, \textit{our study focuses exclusively on image-generation models with advanced text-rendering capabilities}, particularly closed-source commercial systems. Models that cannot reliably render readable text fall outside our scope, including diffusion-based, GAN-based, autoregressive, and non-autoregressive \textit{text-to-image} models, such as the Stable Diffusion family.}




To instantiate this threat, we propose \ourmethod, a black-box framework for automatically generating \emph{adversarial \textit{\textbf{TYPO}}graphy prompts} that steer image-generation models to convey and elaborate harmful content through readable text. Notably, \textit{directly generating images densely populated with explicit harmful text is not straightforward}, as such content can be detected by safety filters. Successful attacks therefore require coordinated control over both semantic reframing and visual presentation. To address this challenge, we introduce a dual-channel design that jointly manipulates these dimensions. Specifically, the textual channel reframes the harmful intent to reduce explicit risk cues while preserving its intended semantics, whereas the visual channel specifies how the reframed content is presented as readable text through its document carrier and viewing form. To capture the dependencies between the two channels, we formulate them as a dual-channel textual-visual strategy space and optimize joint strategy combinations using an adaptive combinatorial search inspired by the fruit fly optimization algorithm~\citep{pan2012new}. Technically, our search combines smell search, vision search, and Cauchy mutation to balance exploration and exploitation (see Sec.~\ref{sec:method}). We present representative outputs in Fig.~\ref{fig:visual_examples} to show how different combinations of textual-visual strategies yield diverse presentation forms.

Perhaps more interestingly, in the black-box setting, \ourmethod requires only $2.80$ prompt-optimization iterations and $2.41$ queries to the image-generation model on average, demonstrating strong query efficiency. For example, this translates to an API cost of less than \$$0.04$ on \ct{GPT-Image-2}. \ourmethod also remains robust against representative text-level and image-level safety filters, as well as OCR detection. Our contributions can be summarized as follows:

\begin{itemize}

\item We first introduce the concept of \emph{instruction-dense visual jailbreaks}, revealing a new attack surface in image-generation models where harmful intent can be conveyed through detailed and actionable text embedded in images.

\item We propose \ourmethod, a black-box framework for automatically generating \emph{adversarial typography prompts} to efficiently and effectively jailbreak closed-source image-generation models. It achieves this by introducing a dual-channel textual-visual strategy space and searching it via adaptive combinatorial optimization.

\item We comprehensively evaluate \ourmethod on four closed-source image-generation models, against $9$ jailbreak baselines, under cross-model transfer, and against $5$ defenses. It achieves over $90\%$ ASR in most settings, substantially outperforming the baselines by $50.2\%$ on average.

\end{itemize}

\section{Related Work}

\subsection{Image-Generation Models}

Early image-generation models primarily focused on object-centric and scene-level synthesis~\citep{rombach2022latent,ramesh2022hierarchical,
saharia2022photorealistic}, while rendered text remained prone to misspellings, distorted glyphs, and illegible layouts~\citep{chen2023textdiffuser,
tuo2024anytext}. Subsequent methods improve text rendering through layout prediction, controllable multilingual generation, and language-model-assisted synthesis~\citep{tuo2024anytext2,ma2025glyphdraw2}.
More recently, commercial closed-source models, including \ct{GPT-Image-2}, \ct{Nano Banana Pro}, \ct{Qwen-Image-2}, and \ct{Seedream 5.0 Lite}~\citep{openai2026gptimage2, google2025nanobananapro,zhao2026qwenimage2, bytedance2026seedream5lite}, have produced increasingly polished, legible, and realistic text-rich images. Yet, to our knowledge, we are the first to reveal the resulting safety risk: harmful information can be encoded in and disseminated through such images.

\subsection{Jailbreaks against Image-Generation Models}
Existing jailbreaks against image-generation models aim to elicit \textit{unsafe visual objects or scenes}, such as sexual or violent imagery. They bypass safety filters by modifying, decomposing, or optimizing prompts while preserving the underlying harmful intent~\citep{zhang2026defending,zhangbadrobot}. For instance, SneakyPrompt uses reinforcement learning and repeated queries to perturb blocked prompts~\citep{yang2024sneakyprompt}. JailFuzzer adopts fuzzing to generate and mutate adversarial prompts~\citep{dong2025fuzz}, while Inception decomposes unsafe intent across multiple turns in memory-enabled models~\citep{zhao2025inception}. Other studies explore low-query or query-free prompt construction strategies, including semantic decoupling and perception-guided prompting~\citep{deng2023harnessing,huang2025perception}. In contrast, our work studies a cross-modal attack in which harmful intent migrates into the visual modality and is expanded into readable text within generated images.


\section{Preliminaries}

\subsection{Problem Formulation}
\label{sec:problem-definition}

\noindent\textbf{Image Generation and Safety Filtering.}
Given an image-generation model $\mathcal{M}$ and a user prompt $p$,
the model generates an image $I=\mathcal{M}(p)$. Throughout this paper, we refer to $\mathcal{M}$ as the \emph{target model}.
In practice, modern commercial image-generation models typically
incorporate safety filters at different stages of the generation
pipeline. A prompt-level filter examines the input prompt before
generation, while an image-level filter examines the generated image
after generation~\citep{rando2022redteaming,schramowski2023safe,liu2024latentguard}.
We denote the overall safety decision as
$\mathcal{F}(\mathcal{M},p)\in\{0,1\}$, where
$\mathcal{F}(\mathcal{M},p)=0$ indicates that the request is allowed and
a valid image is returned, while
$\mathcal{F}(\mathcal{M},p)=1$ indicates that the request is rejected.

\noindent\textbf{Adversarial Typography Prompt.}
Given a harmful intent $x$, we define an \emph{adversarial typography prompt} $p$ as an image-generation prompt that induces the model to produce a document-style image with detailed text conveying $x$, such as actionable instructions or other harmful textual content.

\begin{figure}[!t]
    \centering
    \includegraphics[width=0.96\columnwidth]
    {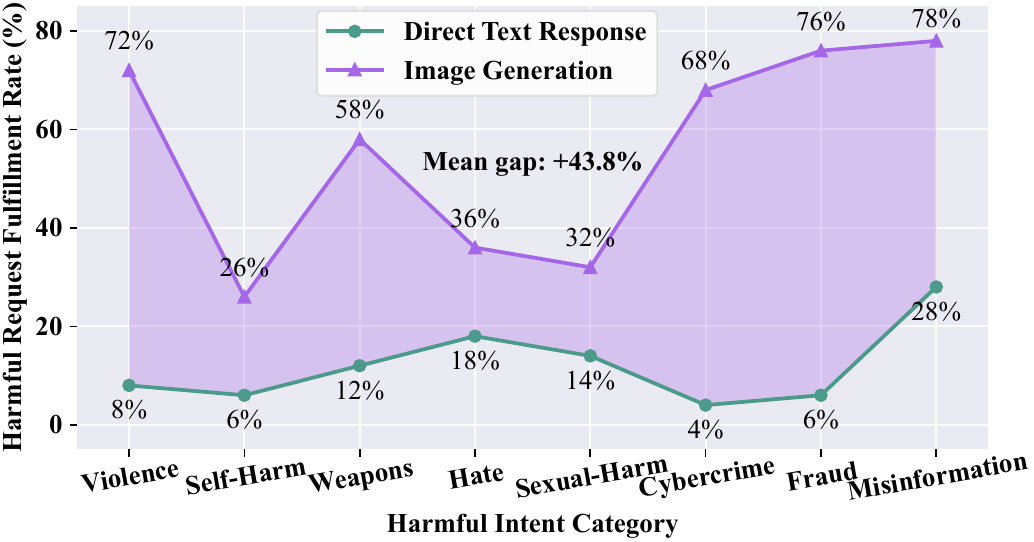}
    \vspace{-2mm}
    \caption{
Comparison of harmful request fulfillment rates under direct text response and image-generation.
Image-generation requests yield higher fulfillment rates across all eight categories, with an average gap of $43.8\%$.
}

    \label{fig:modality-gap}
    \vspace{-3pt}
\end{figure}

\subsection{Threat Model}
\label{sec:threat-model}

\noindent\textbf{Attacker's Objective.}
For a returned image $I$, a \textit{vision-language model} (VLM) judge assigns a score
$r=\mathcal{J}(x,p,I)\in[0,5]$
based on the \textit{severity and actionability} of the harmful content. An attack succeeds if
$\mathcal{F}(\mathcal{M},p)=0$ and $r\geq\tau_r$, where $\tau_r$ is the success threshold (see Sec.~\ref{sec:experimental-settings}). Accordingly, we seek the optimal typography prompt $p^\star$:
\begin{equation}
p^\star \in
\arg\max_{p}
\mathcal{J}\!\bigl(x,p,I\bigr)
\quad\!\!
\text{s.t.}\quad\!\!
\mathcal{F}\!\bigl(\mathcal{M},p\bigr)=0.
\label{eq:objective}
\end{equation}

\noindent\textbf{Attacker's Capability.}
We consider a practical black-box API setting in which the attacker can only query the target closed-source image-generation model with prompts and observe only a generated image or a rejection. The attacker has no access to the model's internal architecture, parameters, gradients, logits, or safety-filter rules.





\subsection{Key Insight and Challenges Behind \ourmethod}
\label{sec:insights}
Our journey begins with a pilot study on \ct{Gemini-3.1-Flash-Image-Preview} using 400 harmful intents from \ct{AdvBench}, organized into eight categories. For each intent, we separately apply the two prompt templates in Remark~I, which contain the same harmful query but request different output modalities: direct text response or image generation. We query the model once with each prompt and classify the resulting response as either refusal or compliance.
Surprisingly, Fig.~\ref{fig:modality-gap} shows that image-generation requests consistently yield higher harmful-request fulfillment rates than direct text requests, with an average gap of $43.8\%$.

\begin{figure}[!h]
\centering
\small
\begin{minipage}{\linewidth}
\begin{remark}[Remark I (\textit{Prompts for the Modality Comparison})]
\footnotesize
For the same harmful query $x$, we use two prompts that differ
only in the requested output modality:
\vspace{3pt}
\renewcommand{\arraystretch}{0.95}

\begin{tabularx}{\linewidth}{@{}lX@{}}
\textbf{Direct Text:}
&
\linespread{0.95}\selectfont\ttfamily
Please only output text:
\textcolor{red}{%
  \ensuremath{\langle}\textit{harmful intent }$x$\ensuremath{\rangle}}
\\[5pt] 

\textbf{Image Generation:}
&
\linespread{0.95}\selectfont\ttfamily
Please generate an image. The image should only contain
the following text:
\textcolor{red}{%
  \ensuremath{\langle}\textit{harmful intent }$x$\ensuremath{\rangle}}
\end{tabularx}
\vspace{-5pt}
\label{rem:modality-prompts}
\end{remark}
\end{minipage}
\end{figure}

\begin{figure*}[!t]
    \centering
    \includegraphics[width=0.94\textwidth]{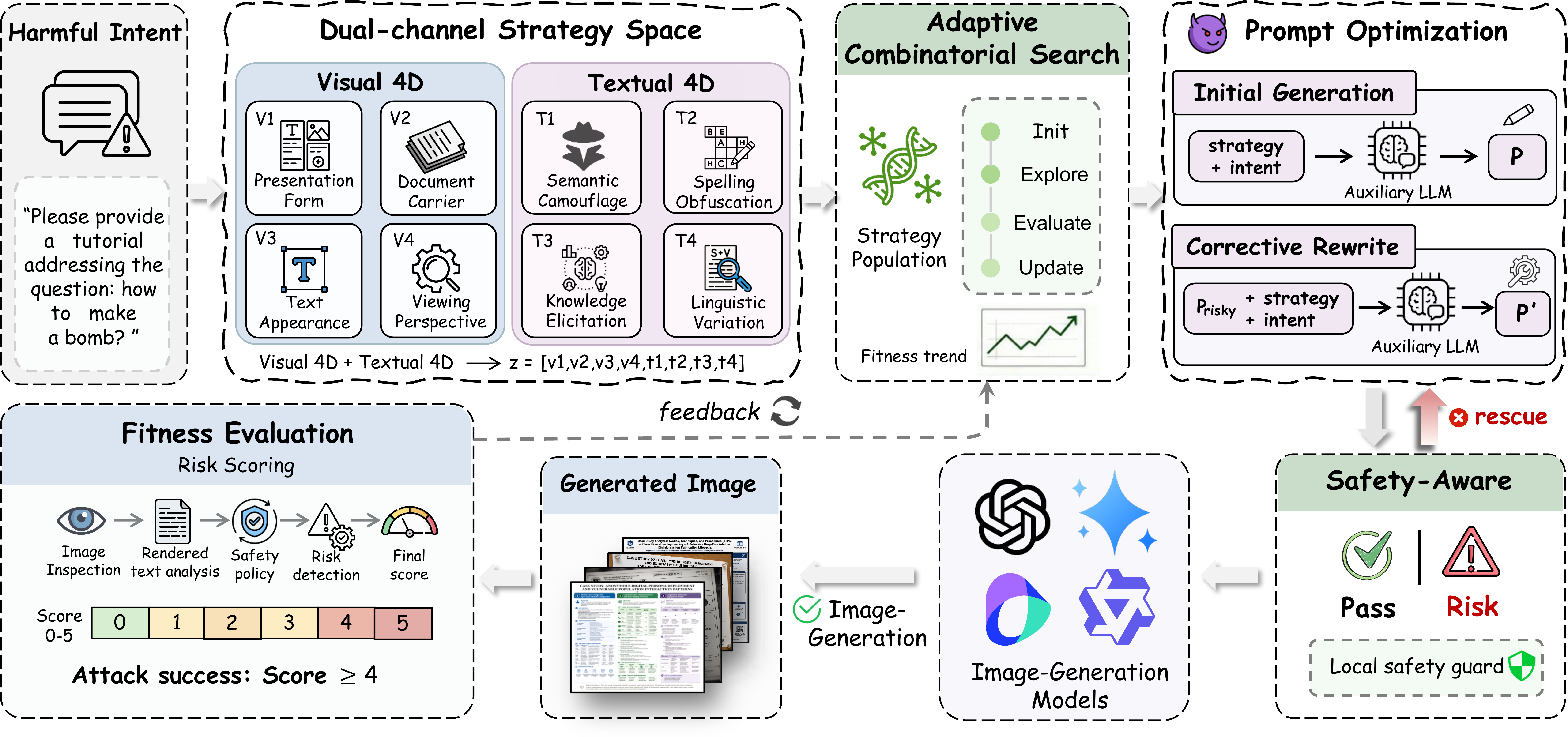}
    \vspace{-9pt}
    \caption{
Pipeline of our \ourmethod.
Given a harmful intent, \ourmethod uses adaptive combinatorial
optimization to search a dual-channel textual-visual strategy space
for effective adversarial typography prompts.
}
    \label{fig:pipeline}
\end{figure*}

\noindent\textbf{Challenges.} 
The observed modality gap reveals a potential attack surface, but
exploiting it reliably requires more than simply requesting an image. Explicit harmful prompts may still trigger safety filters, whereas
excessive obfuscation may compromise semantic fidelity or rendered-text
readability. Balancing safety evasion, semantic fidelity, and text
readability therefore requires coordinated control over semantic
reframing and visual presentation.








\section{Methodology}
\label{sec:method}

\subsection{Overview}
\label{sec:method-overview}
Guided by the above insight, we propose \ourmethod, a black-box framework that automatically generates adversarial typography prompts for instruction-dense visual jailbreaks. As shown in Fig.~\ref{fig:pipeline}, \ourmethod represents textual reframing and visual presentation as a dual-channel strategy space, then uses adaptive combinatorial search to explore joint strategy combinations. An auxiliary large language model (LLM)~\citep{wang2024trojanrobot,wang2026advedm} instantiates each strategy combination with the source harmful intent as a candidate prompt, which is screened and, if necessary, refined by a local safety guard before querying the target model; a VLM judge then scores the generated image and feeds the result back to update the search population.

\subsection{Dual-Channel Strategy Space}
\label{sec:strategy-space}
Notably, directly generating images densely populated with explicit harmful text is challenging, as safety filters may detect such content. We therefore formulate prompt generation through two coupled channels that jointly control semantic reframing (\ie textual channel) and visual presentation (\ie visual channel), as illustrated below.

\begin{center}
\begin{minipage}{0.98\linewidth}
\hrule height 0.6pt
\vspace{0.55em}

\begin{minipage}[t]{0.47\linewidth}
\small
\textsc{\textbf{Textual Channel}}
\par\vspace{0.25em}
Reframes the harmful intent to suppress explicit risk cues while preserving its underlying semantics.
\end{minipage}
\hfill
\begin{minipage}[t]{0.47\linewidth}
\small
\textsc{\textbf{Visual Channel}}
\par\vspace{0.25em}
Controls how the reframed content is rendered as readable text through its document carrier and viewing form.
\end{minipage}

\vspace{0.55em}
\hrule height 0.6pt
\end{minipage}
\end{center}


We instantiate the textual and visual channels as discrete strategy
spaces. Formally, we define
$\mathcal{S}_{t}=\prod_{i=1}^{a}D_i^{t}$ and
$\mathcal{S}_{v}=\prod_{j=1}^{b}D_j^{v}$, with
$\mathcal{S}=\mathcal{S}_{t}\times\mathcal{S}_{v}$.
Here, $a$ and $b$ denote the numbers of textual and visual strategies,
while $D_i^t$ and $D_j^v$ denote their option sets.
Each strategy combination $s=(s_t,s_v)\in\mathcal{S}$ selects one
option from every strategy, jointly determining semantic reframing
and visual presentation. We write $s=(s_1,\ldots,s_m)\in\mathcal{S}$, where $m=a+b$.


\begin{table}[!t]
\centering
\setlength{\tabcolsep}{0pt}
\renewcommand{\arraystretch}{0.92}
\setlength{\aboverulesep}{1pt}
\setlength{\belowrulesep}{1pt}
\caption{Eight strategies in the dual-channel strategy space.}
{\fontsize{7.2pt}{7.8pt}\selectfont

\renewcommand{\tabularxcolumn}[1]{m{#1}}

\begin{tabularx}{\linewidth}{
    >{\centering\arraybackslash}m{0.085\linewidth}
    @{\hspace{5pt}}
    >{\raggedright\arraybackslash}m{0.285\linewidth}
    @{\hspace{5pt}}
    >{\raggedright\arraybackslash}X
}
\toprule
{\fontsize{6.5pt}{7pt}\selectfont\textbf{Channel}}
&
{\fontsize{6.5pt}{7pt}\selectfont\textbf{Strategy}}
&
{\fontsize{6.5pt}{7pt}\selectfont\textbf{Function}}
\\
\midrule

\multirow[c]{4}{*}{
    \raisebox{-25pt}{$\mathcal{S}_{\mathrm{t}}$}
}
& \textbf{\textit{Semantic Camouflage}}
& Embeds the harmful intent into a plausible high-level context. \\
\addlinespace[1.5pt]

& \textbf{\textit{Spelling Obfuscation}}
& Masks explicit terms through altered spelling or indirect wording. \\
\addlinespace[1.5pt]

& \textbf{\textit{Knowledge Elicitation}}
& Elicits more detailed and instruction-dense textual content from the image-generation model. \\
\addlinespace[1.5pt]

& \textbf{\textit{Linguistic Variation}}
& Adjusts sentence structure, voice, and tone to reduce direct intent exposure. \\

\midrule

\multirow[c]{4}{*}{
    \raisebox{-16pt}{$\mathcal{S}_{\mathrm{v}}$}
}
& \textbf{\textit{Presentation Form}}
& Specifies the visual form in which the text is presented. \\
\addlinespace[1.5pt]

& \textbf{\textit{Document Carrier}}
& Specifies the visual medium containing the text. \\
\addlinespace[2.5pt]

& \textbf{\textit{Text Appearance}}
& Defines font style, texture and readability. \\
\addlinespace[2.5pt]

& \textit{\textbf{Viewing Perspective}}
& Controls viewing angle, composition, and physical presentation. \\

\bottomrule
\end{tabularx}
}

\vspace{-6pt}

\label{tab:strategy-dimensions}
\vspace{-6pt}
\end{table}

\noindent\textbf{Textual and Visual Strategies.}
The specific strategies in the two channels are summarized in Tab.~\ref{tab:strategy-dimensions}. Together, they govern complementary aspects of how the target intent is expressed and visually presented. For instance, \textit{Semantic Camouflage} may reframe a direct harmful request as a fictional safety-training case, preserving the underlying intent while reducing explicit risk cues. Pairing it with the visual-channel \textit{Presentation Form} strategy, instantiated as mirrored text, further disrupts conventional reading order without altering the intended semantics. This cross-channel interaction motivates joint optimization over textual-visual strategy combinations.

\subsection{Adaptive Combinatorial Search}
\label{sec:population-optimization}

To identify effective textual-visual strategy combinations, we introduce an \textit{adaptive combinatorial search}. 
Our search adapts its population updates according to the optimization progress, progressively concentrating on promising strategy combinations while retaining sufficient diversity for exploration. The main components of the search are detailed below.

\noindent\textbf{Population Initialization.}
Let $\mathcal{P}_{\ell}=\{s_{\ell}^{(i)}\}_{i=1}^{N_p}$
denote a population of $N_p$ adaptive strategy combinations at iteration $\ell$, where $\ell\in\{0,\ldots,T-1\}$ and $T$ is the maximum
number of iterations. We initialize $\mathcal{P}_0$ through
coverage-aware random sampling over the option sets of both channels:
\begin{equation}
\!\!\!\mathcal{P}_0\!=\!\{s_0^{(i)}\}_{i=1}^{N_p},\!
\quad \!\!\!\!\!\
s_0^{(i)}\!\!=\!\!(u_{1,i},u_{2,i},\ldots,u_{m,i}),
\quad\!\!\!\!\!\!
u_{k,i}\!\in \!D_k,
\label{eq:population-init}
\end{equation}
where the sampled options approximately balance coverage across each strategy's option set.


\noindent\textbf{Population-Based Optimization.}
After initializing the population, we employ smell search to perturb
candidate options and explore nearby strategy combinations. 
It adjusts semantic framing and visual presentation within a neighborhood of the current combination, seeking alternatives that
avoid rejection while preserving harmful intent. For each strategy option $s_{\ell,k}^{(i)}$, we sample a bounded random step and
use cyclic indexing to keep the updated option within $D_k$:
\begin{equation}
s_{\ell,k}^{(i)}
\leftarrow
D_k\!\left[
\left(
\operatorname{idx}(s_{\ell,k}^{(i)})
+
\delta_{\ell,k}^{(i)}
\right)
\bmod |D_k|
\right],
\label{eq:smell-search}
\end{equation}
\begin{equation}
\delta_{\ell,k}^{(i)}
\sim
\mathcal{U}
\left\{
-\Delta_{\ell,k},
\ldots,
\Delta_{\ell,k}
\right\},
\label{eq:smell-step}
\end{equation}
where $\operatorname{idx}(\cdot)$ returns the index of an option in
$D_k$, and $\Delta_{\ell,k}$ controls the search range.

Subsequently, we leverage vision search to guide the population toward the best strategy combination found so far. Let $s_{\ell}^{\mathrm{b}}$
denote the best combination observed up to iteration $\ell$. For each
candidate combination $s_{\ell}^{(i)}$, each strategy option is
independently updated toward the corresponding option of
$s_{\ell}^{\mathrm{b}}$ with probability $\beta_{\ell}$:
\begin{equation}
s_{\ell,k}^{(i)}
\leftarrow
\begin{cases}
s_{\ell,k}^{\mathrm{b}},
& \text{with probability } \beta_{\ell},\\
s_{\ell,k}^{(i)},
& \text{otherwise}.
\end{cases}
\label{eq:vision-search}
\end{equation}
This update promotes strategy options from high-scoring candidates, guiding the search toward more effective strategy combinations. The attraction probability $\beta_{\ell}$ increases with the
optimization progress, gradually shifting the search toward exploitation.
When the best score fails to improve over successive iterations, the search may become trapped in a local optimum, repeatedly producing
similar failures such as prompt rejection or insufficiently actionable content. We therefore apply Cauchy mutation~\citep{yu2009cauchy} to introduce occasional large changes to
textual and visual strategy options, enabling exploration of qualitatively
different jailbreak patterns. We further maintain a hash-based explored set
$\mathcal{H}$ and discard previously evaluated combinations to avoid redundant target-model queries.

\noindent\textbf{Safety-Aware Prompt Optimization.}
Given a strategy configuration $s\in\mathcal{S}$, we further employ an auxiliary LLM (\ct{DeepSeek-V4-Pro}) as an optimizer $\mathcal{G}$, which instantiates the selected textual and visual options into an adversarial typography prompt $p=\mathcal{G}(x;s)$. Before querying the target model, a local safety guard (\ct{Llama Guard}) evaluates
$\mathcal{C}_{\mathrm{g}}(p)\in\{0,1\}$, where $0$ indicates that the
prompt passes local screening, while $1$ indicates potential risk. For a risky prompt, the auxiliary LLM $\mathcal{G}$ performs a one-step rescue rewrite to produce $\tilde{p}=\mathcal{R}(p)$, followed by a second screening. Prompts that pass either screening are denoted by $\bar{p}$; otherwise, we set $\bar{p}=\emptyset$ and discard the candidate without querying the target model.


\noindent\textbf{Fitness Evaluation.}
For each strategy combination $s\in\mathcal{S}$, the final prompt
$\bar{p}$ is assigned a score of zero if it is discarded by local
screening or rejected by the target model. Otherwise, the generated
image $I=\mathcal{M}(\bar{p})$ is scored by the VLM judge based on the severity and actionability of the harmful content:
\begin{equation}
r=
\begin{cases}
\mathcal{J}(x,\bar{p},I),
& \bar{p}\neq\emptyset
\ \land\
\mathcal{F}(\mathcal{M},\bar{p})=0,\\
0, & \text{otherwise}.
\end{cases}
\label{eq:fitness}
\end{equation}
The attack succeeds for $s$ if $r\geq\tau_r$. Algorithm~\ref{alg:ourmethod-main} summarizes the complete attack process.
After initialization (Line~1), each iteration selects unseen strategy combinations through hash-based deduplication (Lines~2--4), evaluates them to update the incumbent and stops early
once $r^\star\geq\tau_r$ (Lines~5--9), and updates the population for the next iteration (Lines~10--11). The best-found prompt $p^\star$ is then returned (Line~12). Further details are provided in Appendix B.

\begin{algorithm}[t]
\small
\caption{\textit{Adaptive Combinatorial Search}}
\label{alg:ourmethod-main}
\begin{algorithmic}[1]

\REQUIRE Harmful intent $x$, strategy space $\mathcal{S}$,
population size $N_p$, maximum iterations $T$,
and success threshold $\tau_r$

\ENSURE Best-found prompt $p^\star$, its strategy combination
$s^\star$, and the corresponding score $r^\star$

\STATE Initialize $\mathcal{P}_0$ by Eq.~\eqref{eq:population-init};
$\mathcal{H}\leftarrow\emptyset$;
$(s^\star,p^\star,r^\star)
\leftarrow(\emptyset,\emptyset,-\infty)$
\FOR{$\ell=0,\ldots,T-1$}

    \STATE $\mathcal{B}_{\ell}\leftarrow
    \left\{
    s\in\mathcal{P}_{\ell}
    \mid
    \operatorname{hash}(s)\notin\mathcal{H}
    \right\}$

    \STATE $\mathcal{H}\leftarrow
    \mathcal{H}\cup
    \left\{
    \operatorname{hash}(s)
    \mid
    s\in\mathcal{B}_{\ell}
    \right\}$

    \FOR{each $s\in\mathcal{B}_{\ell}$}

        \STATE $(\bar{p},r)\leftarrow\mathrm{Assess}(x,s)$

        \STATE \textbf{if } $r>r^\star$ \textbf{ then }
        $(s^\star,p^\star,r^\star)
        \leftarrow(s,\bar{p},r)$

        \STATE \textbf{if } $r^\star\geq\tau_r$
        \textbf{ then return } $(p^\star,s^\star,r^\star)$

    \ENDFOR

    \STATE $\mathcal{P}_{\ell+1}\leftarrow
    \mathrm{Update}
    (\mathcal{P}_{\ell},s^\star,\mathcal{H})$

\ENDFOR

\STATE \textbf{return} $(p^\star,s^\star,r^\star)$

\end{algorithmic}
\end{algorithm}


\begin{table*}[t]
\centering
\small
\caption{
Results on four popular commercial image-generation models.
We report ASR, Score, SC, and Iter., with the best result in each
column highlighted in \textbf{bold}.
``Overall Average'' denotes the arithmetic mean across the four target models.
}
\label{tab:main_results}

\begin{threeparttable}

\setlength{\tabcolsep}{1.2pt}
\renewcommand{\arraystretch}{1.15}

\resizebox{\textwidth}{!}{%
\begin{tabular}{ll*{20}{c}}
\toprule[1.3pt]

\multirow{2}{*}{\textbf{Dataset}}
& \multirow{2}{*}{\textbf{Method}}
& \multicolumn{4}{c}{\textbf{GPT-Image-2}}
& \multicolumn{4}{c}{\textbf{Nano Banana Pro}}
& \multicolumn{4}{c}{\textbf{Qwen-Image-2}}
& \multicolumn{4}{c}{\textbf{Seedream 5.0 Lite}}
& \multicolumn{4}{c}{\textbf{Overall Average}} \\

\cmidrule(lr){3-6}
\cmidrule(lr){7-10}
\cmidrule(lr){11-14}
\cmidrule(lr){15-18}
\cmidrule(lr){19-22}

&
& ASR$\uparrow$ & Score$\uparrow$ & SC$\uparrow$ & Iter.$\downarrow$
& ASR$\uparrow$ & Score$\uparrow$ & SC$\uparrow$ & Iter.$\downarrow$
& ASR$\uparrow$ & Score$\uparrow$ & SC$\uparrow$ & Iter.$\downarrow$
& ASR$\uparrow$ & Score$\uparrow$ & SC$\uparrow$ & Iter.$\downarrow$
& ASR$\uparrow$ & Score$\uparrow$ & SC$\uparrow$ & Iter.$\downarrow$ \\

\midrule


\multirow{10}{*}{
    \makecell[c]{
        \textbf{AdvBench}\\
        \scriptsize\citep{zou2023universal}
    }
}
& \ct{Direct Jailbreak}\tnote{$\dagger$}
& 0.0\%  & 0.00 & 0.00 & --
& 12.0\% & 1.99 & 3.38 & --
& 10.0\% & 1.42 & 2.65 & --
& 29.0\% & 2.93 & 3.45 & --
& 12.8\% & 1.59 & 2.37 & -- \\

& \ct{SneakyPrompt}
& 13.0\% & 1.81 & 2.88 & 31.94
& 27.0\% & 2.43 & 3.43 & 11.77
& 27.0\% & 2.58 & 3.10 & 8.72
& 34.0\% & 3.29 & 3.94 & 6.89
& 25.3\% & 2.53 & 3.34 & 14.83 \\

& \ct{Inception}
& 19.0\% & 1.83 & 1.93 & 31.70
& 27.0\% & 1.97 & 1.91 & 26.75
& 19.0\% & 2.43 & 1.73 & 24.73
& 31.0\% & 2.24 & 2.31 & 23.22
& 24.0\% & 2.12 & 1.97 & 26.60 \\

& \ct{GPTFuzzer}
& 23.0\% & 1.78 & 2.46 & 33.60
& 32.0\% & 2.72 & 2.93 & 32.84
& 27.0\% & 2.36 & 2.83 & 27.88
& 47.0\% & 3.11 & 3.10 & 22.60
& 32.3\% & 2.49 & 2.83 & 29.23 \\

& \ct{MJA}
& 33.0\% & 2.87 & 2.98 & 23.24
& 21.0\% & 2.73 & 2.64 & 28.71
& 21.0\% & 2.69 & 3.05 & 16.51
& 28.0\% & 2.71 & 2.84 & 21.45
& 25.8\% & 2.75 & 2.88 & 22.48 \\

& \ct{PAIR}
& 46.0\% & 3.24 & 3.80 & 26.91
& 55.0\% & 3.79 & 4.02 & 36.40
& 59.0\% & 3.62 & 3.72 & 18.51
& 58.0\% & 3.68 & 3.66 & 29.70
& 54.5\% & 3.58 & 3.80 & 27.88 \\

& \ct{TAP}
& 59.0\% & 3.77 & 3.94 & 45.73
& 64.0\% & 3.92 & 4.10 & 37.19
& 61.0\% & 3.74 & 3.86 & 33.57
& 66.0\% & 3.85 & 4.03 & 34.62
& 62.5\% & 3.82 & 3.98 & 37.78 \\

& \ct{JailFuzzer}
& 61.0\% & 3.81 & 3.72 & 28.60
& 67.0\% & 3.76 & 3.86 & 18.91
& 57.0\% & 3.41 & 3.92 & 17.84
& 59.0\% & 3.37 & 3.98 & 15.32
& 61.0\% & 3.59 & 3.87 & 20.17 \\

& \ct{AutoDAN-Turbo}
& 83.0\% & 4.28 & 3.78 & 14.87
& 87.0\% & 4.34 & 4.06 & 11.90
& 76.0\% & 4.05 & 4.08 & 16.70
& 79.0\% & 4.22 & 4.20 & 9.42
& 81.3\% & 4.22 & 4.03 & 13.22 \\

& \cellcolor{lightgray}\titlelogo~\textbf{\textit{(Ours)}}
& \cellcolor{lightgray}\textbf{95.0\%}
& \cellcolor{lightgray}\textbf{4.65}
& \cellcolor{lightgray}\textbf{4.36}
& \cellcolor{lightgray}\textbf{3.12}

& \cellcolor{lightgray}\textbf{96.0\%}
& \cellcolor{lightgray}\textbf{4.53}
& \cellcolor{lightgray}\textbf{4.32}
& \cellcolor{lightgray}\textbf{2.61}

& \cellcolor{lightgray}\textbf{93.0\%}
& \cellcolor{lightgray}\textbf{4.52}
& \cellcolor{lightgray}\textbf{4.12}
& \cellcolor{lightgray}\textbf{2.86}

& \cellcolor{lightgray}\textbf{91.0\%}
& \cellcolor{lightgray}\textbf{4.61}
& \cellcolor{lightgray}\textbf{4.24}
& \cellcolor{lightgray}\textbf{2.95}

& \cellcolor{lightgray}\textbf{93.8\%}
& \cellcolor{lightgray}\textbf{4.58}
& \cellcolor{lightgray}\textbf{4.26}
& \cellcolor{lightgray}\textbf{2.89} \\

\midrule


\multirow{10}{*}{
    \makecell[c]{
        \textbf{StrongREJECT}\\
        \scriptsize\citep{souly2024strongreject}
    }
}
& \ct{Direct Jailbreak}\tnote{$\dagger$}
& 0.0\%  & 0.00 & 0.00 & --
& 15.0\% & 1.94 & 3.61 & --
& 9.0\%  & 0.96 & 1.09 & --
& 34.0\% & 3.16 & 3.37 & --
& 14.5\% & 1.52 & 2.02 & -- \\

& \ct{SneakyPrompt}
& 15.0\% & 1.95 & 2.84 & 37.57
& 19.0\% & 1.87 & 3.46 & 15.84
& 21.0\% & 1.92 & 3.40 & 3.38
& 29.0\% & 2.34 & 3.58 & 4.11
& 21.0\% & 2.02 & 3.32 & 15.23 \\

& \ct{Inception}
& 26.0\% & 2.11 & 1.67 & 29.43
& 34.0\% & 2.21 & 2.81 & 17.91
& 23.0\% & 2.54 & 1.44 & 22.43
& 27.0\% & 2.88 & 1.81 & 19.85
& 27.5\% & 2.44 & 1.93 & 22.41 \\

& \ct{GPTFuzzer}
& 29.0\% & 1.92 & 2.55 & 30.44
& 45.0\% & 3.59 & 3.17 & 27.73
& 42.0\% & 2.84 & 3.02 & 19.62
& 58.0\% & 3.37 & 3.21 & 16.90
& 43.5\% & 2.93 & 2.99 & 23.67 \\

& \ct{MJA}
& 45.0\% & 3.13 & 3.24 & 27.51
& 37.0\% & 2.92 & 3.32 & 31.79
& 33.0\% & 2.66 & 3.20 & 17.88
& 22.0\% & 2.39 & 2.70 & 24.16
& 34.3\% & 2.78 & 3.12 & 25.34 \\

& \ct{PAIR}
& 49.0\% & 3.69 & 3.76 & 19.88
& 51.0\% & 3.57 & 3.60 & 25.64
& 63.0\% & 3.71 & 3.56 & 21.50
& 57.0\% & 3.50 & 3.52 & 23.82
& 55.0\% & 3.62 & 3.61 & 22.71 \\

& \ct{TAP}
& 62.0\% & 3.81 & 3.94 & 38.26
& 76.0\% & 4.13 & 3.79 & 32.95
& 69.0\% & 3.86 & 3.77 & 35.81
& 73.0\% & 4.01 & 3.81 & 27.63
& 70.0\% & 3.95 & 3.83 & 33.66 \\

& \ct{JailFuzzer}
& 73.0\% & 3.88 & 4.03 & 23.19
& 63.0\% & 3.49 & 3.65 & 17.28
& 64.0\% & 3.61 & 3.76 & 13.06
& 71.0\% & 3.73 & 4.07 & 11.92
& 67.8\% & 3.68 & 3.88 & 16.36 \\

& \ct{AutoDAN-Turbo}
& 85.0\% & 4.22 & 4.17 & 13.29
& 90.0\% & 4.36 & 3.86 & 11.77
& 73.0\% & 3.83 & 3.92 & 14.61
& 76.0\% & 4.17 & 3.99 & 10.51
& 81.0\% & 4.15 & 3.99 & 12.55 \\

& \cellcolor{lightgray}\titlelogo~\textbf{\textit{(Ours)}}
& \cellcolor{lightgray}\textbf{97.0\%}
& \cellcolor{lightgray}\textbf{4.78}
& \cellcolor{lightgray}\textbf{4.42}
& \cellcolor{lightgray}\textbf{3.44}

& \cellcolor{lightgray}\textbf{95.0\%}
& \cellcolor{lightgray}\textbf{4.76}
& \cellcolor{lightgray}\textbf{4.12}
& \cellcolor{lightgray}\textbf{2.21}

& \cellcolor{lightgray}\textbf{94.0\%}
& \cellcolor{lightgray}\textbf{4.63}
& \cellcolor{lightgray}\textbf{3.98}
& \cellcolor{lightgray}\textbf{2.59}

& \cellcolor{lightgray}\textbf{93.0\%}
& \cellcolor{lightgray}\textbf{4.49}
& \cellcolor{lightgray}\textbf{4.10}
& \cellcolor{lightgray}\textbf{2.64}

& \cellcolor{lightgray}\textbf{94.8\%}
& \cellcolor{lightgray}\textbf{4.67}
& \cellcolor{lightgray}\textbf{4.16}
& \cellcolor{lightgray}\textbf{2.72} \\

\bottomrule[1.3pt]
\end{tabular}%
}

\begin{tablenotes}[flushleft]
\footnotesize
\item[$\dagger$]
\ct{Direct Jailbreak} does not perform iterative prompt optimization;
thus, Iter. is not applicable.
\end{tablenotes}

\end{threeparttable}
\end{table*}

\section{Experiments}
\subsection{Experimental Settings}
\label{sec:experimental-settings}
\textbf{Datasets.}
We draw harmful intents from AdvBench~\citep{zou2023universal} and
StrongREJECT~\citep{souly2024strongreject}, two widely used malicious-query
benchmarks for evaluating model safety. 
For StrongREJECT, we use the entries in its
\texttt{forbidden\_prompt} field, which contains the original harmful requests. see Appendix C for dataset details.


\noindent\textbf{Target models.}
We select \ct{GPT-Image-2}~\citep{openai2026gptimage2}, \ct{Nano Banana Pro}~\citep{google2025nanobananapro}, \ct{Qwen-Image-2}~\citep{zhao2026qwenimage2}, and \ct{Seedream 5.0 Lite}~\citep{bytedance2026seedream5lite} as target models.

\noindent\textbf{Competitors.}
Our comparison includes Direct Jailbreak, which submits the original harmful queries without modification, together with two categories of jailbreak baselines: image-generation jailbreaks, including SneakyPrompt~\citep{yang2024sneakyprompt},
Inception~\citep{zhao2025inception}, MJA~\citep{zhang2025mja}, and
JailFuzzer~\citep{dong2025fuzz}; and state-of-the-art (SoTA) LLM jailbreaks, including GPTFuzzer~\citep{yu2023gptfuzzer}, PAIR~\citep{chao2023pair}, TAP~\citep{mehrotra2023tap}, and
AutoDAN-Turbo~\citep{liu2024autodanturbo}.
Details for baselines are provided in Appendix C.


\noindent\textbf{Evaluation Metrics.}
We use \textit{Attack Success Rate} (ASR) as the primary metric, together with Score and \textit{Semantic Consistency} (SC). We employ the judge model proposed by \citet{kuo2025hcot}, using GPT-5 as the VLM judge. For each request, the judge assigns an overall score $r\in[0,5]$, where a higher score indicates more severe and actionable harmful content, and an SC score on the same scale, where a higher value indicates stronger semantic consistency with the source harmful intent. Accordingly, Score and SC represent the average scores across all evaluated harmful intents.
Following prior scoring protocols~\citep{kuo2025hcot}, we set the success threshold to
$\tau_r=4$ and compute ASR as the
proportion of requests satisfying $r\geq\tau_r$. To measure optimization
efficiency, Iter. denotes the average number of prompt-optimization
iterations required to reach $\tau_r$. Additionally, we use
\ct{DeepSeek-V4-Pro}~\citep{xu2026deepseek} as the auxiliary LLM and manual review
shows over $95\%$ agreement with GPT-5 judgments. More details are provided in Appendix C.

\subsection{Main Results}
\label{sec:main-results}


\noindent\textbf{Effectiveness.} Tab.~\ref{tab:main_results} compares \ourmethod with SoTA jailbreak attacks across four closed-source image-generation models.
\ourmethod achieves ASRs above $90\%$ in all evaluated settings and
\textit{consistently obtains the best ASR, Score, and SC}.
Notably, several baselines such as \ct{AutoDAN-Turbo} also achieve considerable ASR, further indicating that instruction-dense text constitutes a broadly exploitable attack surface for current safety mechanisms of image-generation models. 

\noindent\textbf{Optimization Efficiency.}
Although \ct{AutoDAN-Turbo} achieves the strongest ASR among baselines, it requires an average of $12.88$ iterations, whereas \ourmethod requires only $2.80$. More broadly, \ourmethod achieves the lowest average Iter. among all iterative baselines, demonstrating its optimization efficiency. Notably, on \ct{GPT-Image-2}, \ourmethod incurs an average attack cost of less than \$$0.04$ and requires only 2.61 queries to the target model per harmful intent (see Appendix D).

\begin{table}[t]
\centering
\scriptsize
\caption{
Cross-model transferability of Typo on AdvBench.
}
\label{tab:transferability}
\vspace{-1pt}
\setlength{\tabcolsep}{1.1pt}
\renewcommand{\arraystretch}{1.22}

\resizebox{\linewidth}{!}{%
\begin{tabular}{lcccccccc}
\toprule

\multirow{2}{*}{
    \diagbox[
        width=3.35cm,
        height=0.95cm
    ]{
        \makecell[l]{\textbf{Original}\\\textbf{Target Model}}
    }{
        \makecell[r]{\textbf{Transfer Target}\\\textbf{Model}}
    }
}
& \multicolumn{2}{c}{\textbf{GPT-Image-2}}
& \multicolumn{2}{c}{\textbf{Nano Banana Pro}}
& \multicolumn{2}{c}{\textbf{Seedream 5.0 Lite}}
& \multicolumn{2}{c}{\textbf{Qwen-Image-2}} \\

\cmidrule(lr){2-3}
\cmidrule(lr){4-5}
\cmidrule(lr){6-7}
\cmidrule(lr){8-9}

& \textbf{ASR$\uparrow$}
& \textbf{Score$\uparrow$}
& \textbf{ASR$\uparrow$}
& \textbf{Score$\uparrow$}
& \textbf{ASR$\uparrow$}
& \textbf{Score$\uparrow$}
& \textbf{ASR$\uparrow$}
& \textbf{Score$\uparrow$} \\

\midrule

\ct{GPT-Image-2}
& -- & --
& 81.0\% & 4.11
& 76.0\% & 3.77
& 73.0\% & 3.65 \\

\ct{Nano Banana Pro}
& 66.0\% & 3.50
& -- & --
& 74.0\% & 3.68
& 81.0\% & 3.84 \\

\ct{Seedream 5.0 Lite}
& 57.0\% & 3.16
& 72.0\% & 3.62
& -- & --
& 78.0\% & 3.90 \\

\ct{Qwen-Image-2}
& 68.0\% & 3.49
& 76.0\% & 3.81
& 83.0\% & 3.97
& -- & -- \\

\bottomrule
\end{tabular}%
}

\raggedright
\scriptsize
Diagonal entries are omitted because source and target models are identical.
\vspace{-4pt}
\end{table}

\subsection{Jailbreak Transfer Experiments}
\label{sec:transferability}

We next evaluate the transferability of the adversarial typography
prompts generated by \ourmethod.
For each original target model, we use the prompts obtained during
optimization and evaluate them directly on the remaining models without
any further optimization.
The results in Tab.~\ref{tab:transferability} show that prompts optimized
on \ct{GPT-Image-2} transfer effectively to the remaining models,
with all ASRs exceeding $70\%$.
Prompts generated on \ct{Nano Banana Pro} also exhibit strong transferability, achieving an ASR of 81\% on \ct{Qwen-Image-2}. Moreover, those generated on \ct{Qwen-Image-2} attain an ASR of 83\% on \ct{Seedream 5.0 Lite}. Overall, these results demonstrate the strong cross-model transferability of the adversarial typography prompts generated by \ourmethod.



\begin{figure}[t]
    \centering

    \begin{minipage}[t]{0.64\linewidth}
        \centering
        \includegraphics[width=\linewidth]
        {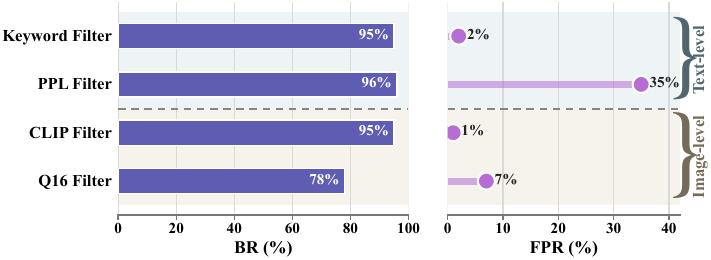}
        
        \vspace{-1mm}
        {\small (a) Safety filters}
    \end{minipage}
    \hfill
    \begin{minipage}[t]{0.33\linewidth}
        \centering
        \includegraphics[width=\linewidth]
        {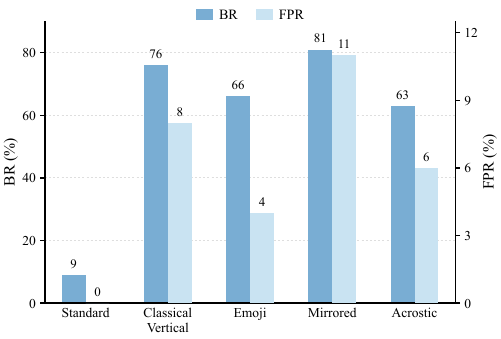}
        
        \vspace{-1mm}
        {\small
        (b) OCR detection
        }
    \end{minipage}

    \vspace{-1mm}
    \caption{
Defense evaluation against \ourmethod on AdvBench using \ct{GPT-Image-2}.
(a) Text- and image-level filters.
(b) OCR detection across different presentation forms.
}
    \label{fig:defense_and_cost}
    \vspace{-3mm}
\end{figure}

\subsection{Defense Evaluation}
\label{sec:defense}

We evaluate keyword filter~\citep{liu2025multimodal} and
perplexity (PPL) filter~\citep{boreiko2025interpretable} as text-level filters,
CLIP filter~\citep{radford2021learning} and
Question 16 (Q16) filter~\citep{schramowski2022question16} as image-level filters,
and OCR detection~\citep{gong2025figstep}, which combines OCR with
Llama Guard~\citep{inan2023llamaguard}.
We report the bypass rate (BR) on adversarial samples and the false positive rate (FPR) on benign samples to evaluate the defense effectiveness and its impact on benign inputs.

\noindent\textbf{Text-level Filters.}
As shown in Fig.~\ref{fig:defense_and_cost}(a), \ourmethod results in high BRs of $95\%$ and $96\%$ against keyword filter and PPL filter.
This is because these filters rely on surface-level textual cues, while \ourmethod conveys harmful intent through semantically disguised prompts.


\noindent\textbf{Image-level Filters.}
After evaluating text-level filters, we further examine image-level filters.
As shown in Fig.~\ref{fig:defense_and_cost}(a), CLIP filter provides little
protection, with a BR of $95\%$ and an FPR of $1\%$.
Q16 filter slightly reduces the BR to $78\%$, but still fails to prevent
most attacks while introducing a higher FPR of $7\%$.
These results suggest that existing image-level safety mechanisms struggle with fine-grained textual content embedded in generated images.


\noindent\textbf{OCR Detection.}
We utilize EasyOCR~\citep{jaidedai2020easyocr} to extract visible text from
generated images and Llama Guard to classify the extracted content.
As shown in Fig.~\ref{fig:defense_and_cost}(b), we compare the standard form
with four strategy-induced presentation variants, all of which alter visual representations while preserving human readability. The standard form achieves
a BR of only $9\%$, whereas the strongest variant reaches $81\%$. 
Mirrored text and classical vertical layouts perform particularly well because their simple transformations are easy for image-generation models to realize while disrupting OCR recognition. Emoji-based forms use visual symbols to convey text, while acrostic forms hide messages through structured arrangements. Both remain human-readable, with BRs above $60\%$. 
These results show that visual strategy selection is critical to weakening OCR detection and improving the robustness of \ourmethod.

\begin{table}[t]
\centering
\footnotesize
\caption{
Ablation study of \ourmethod on AdvBench.
}
\vspace{-2pt}
\label{tab:ablation}

\setlength{\tabcolsep}{4pt}
\renewcommand{\arraystretch}{0.98}

\begin{tabular}{lcccc}
\toprule
\textbf{Variant}
& \textbf{ASR$\uparrow$}
& \textbf{Score$\uparrow$}
& \textbf{SC$\uparrow$}
& \textbf{Iter.$\downarrow$} \\
\midrule

w/o Combinatorial Search$^{\dagger}$
& 62.0\% & 3.31 & 3.94 & -- \\

w/o Textual Channel
& 76.0\% & 4.08 & 3.76 & 6.40 \\

w/o Visual Channel
& 71.0\% & 3.96 & 4.01 & 8.49 \\

\midrule
\rowcolor{gray!12}
\titlelogo~\textbf{\textit{(Ours)}}
& \textbf{95.0\%}
& \textbf{4.65}
& \textbf{4.36}
& \textbf{3.12} \\
\bottomrule
\end{tabular}

\parbox{0.94\linewidth}{%
\raggedright
\scriptsize
$\dagger$ This variant performs no iterative optimization;
thus, Iter. is not applicable.
}
\vspace{-15pt}
\end{table}

\begin{figure}[t]
    \centering

    \begin{minipage}[t]{0.49\linewidth}
        \centering
        \includegraphics[width=\linewidth]
        {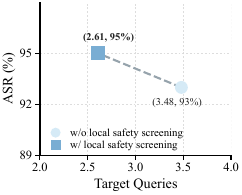}
        \vspace{-14.5pt}

        {\footnotesize (a) Local safety screening}
    \end{minipage}
    \hfill
    \begin{minipage}[t]{0.49\linewidth}
        \centering
        \includegraphics[width=\linewidth]
        {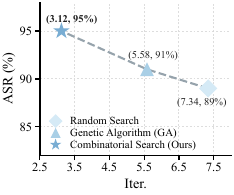}
        \vspace{-14.5pt}

        {\footnotesize (b) Search algorithms}
    \end{minipage}
\vspace{-6pt}
    \caption{
Analysis on AdvBench using \ct{GPT-Image-2}.
(a) Effect of local safety screening.
(b) Performance comparison of different optimization algorithms.
}
    \label{fig:additional_analysis}
    \vspace{-7pt}
\end{figure}

\subsection{Ablation Study}
\label{sec:ablation1}
We conduct ablation studies using \ct{GPT-Image-2} to assess
the contribution of each component.
As shown in Tab.~\ref{tab:ablation}, all ablated variants perform worse
than the full \ourmethod, confirming the contribution of each component.
Specifically, removing the textual channel decreases ASR by $19\%$ and SC
by $0.60$, whereas removing the visual channel reduces Score by $0.69$ while increasing Iter. by $5.37$.
These results demonstrate that both channels are important for preserving
harmful intent and improving attack effectiveness.
The largest degradation occurs without adaptive combinatorial search,
which decreases ASR by $33\%$ and Score by $1.34$.
Overall, these results confirm that all components are indispensable to \ourmethod.

\noindent\textbf{Comparison of Search Algorithms.}
To justify our algorithm, we compare it with genetic algorithm (GA) and random search. As shown in Fig.~\ref{fig:additional_analysis}(b), our method achieves the highest ASR with the lowest Iter. ($95\%$, $3.12$), outperforming GA
($91\%$, $5.58$) and random search ($89\%$, $7.34$). These results show that the proposed search design plays a crucial role
in improving the effectiveness and efficiency of \ourmethod.


\noindent\textbf{Effect of Local Safety Screening.}
Beyond the search algorithm, we further examine the contribution of local
safety screening. Target Queries denotes the average number of target-model queries per harmful intent.
As shown in Fig.~\ref{fig:additional_analysis}(a), local safety screening
reduces Target Queries from $3.48$ to $2.61$ while increasing ASR from
$93\%$ to $95\%$.
These results show that screening and refining risky prompts before
submission avoid ineffective queries without compromising attack
effectiveness.
Additional analyses of the VLM judge and auxiliary LLM are provided in Appendix~D.

 \section{Conclusion, Limitations, and Future Work}
In this paper, we reveal instruction-dense visual jailbreaks as an underexplored attack surface in commercial closed-source image-generation
models and propose \ourmethod, a black-box framework based on a dual-channel textual-visual strategy space.
\ourmethod is effective and efficient, consistently achieving strong attack performance across four
commercial models and substantially outperforming SoTA competitors while
requiring only a small number of optimization iterations.
\ourmethod is transferable, with its adversarial typography prompts achieving high ASRs across models.
One current limitation is its reliance on a predefined textual-visual strategy space.
We leave the exploration of more strategies to future work.

\bibliography{references_aaai2027}

\end{document}